\documentclass[%
 reprint,
 amsmath,amssymb,
 aps,
]{revtex4-2}

\usepackage{graphicx}% Include figure files
\usepackage{dcolumn}% Align table columns on decimal point
\usepackage{bm}% bold math
\usepackage{physics}
\usepackage{tikz}
\usepackage{algorithm}
\usepackage{algpseudocode}
\usepackage{tabularx}
\usepackage{booktabs}
\usepackage{subcaption}% Include figure files
\usepackage{hyperref}% add hypertext capabilities
\usepackage{adjustbox}
\usepackage{color}
\usepackage{tikz}
\begin{document}

\preprint{APS/123-QED}

\title{Learning magic in the Schwinger model}% Force line breaks with \\

\author{Samuel Crew}
\email{samuel.c.crew@gmail.com}
\affiliation{Department of Physics, National Tsing Hua University, Hsinchu 30013, Taiwan}

\author{Hsueh Hao Lu}
\affiliation{Department of Physics, National Tsing Hua University, Hsinchu 30013, Taiwan}

\date{\today}% It is always \today, today,

\begin{abstract}
We demonstrate the use of variational neural network quantum states to study non-stabilizerness in qubit-regularised quantum field theory. Applying the methodology recently introduced by Sinibaldi \textit{et al.}, we numerically compute the stabilizer Rényi entropy of ground states of the Schwinger model with a topological term. We examine how the magic content of these states depends on the separation between external probe charges, providing insight into the classical hardness of simulating gauge theories with non-trivial infrared structure.
\end{abstract}

\maketitle

%\tableofcontents

\section{Introduction}

Entanglement has long been recognised as a key resource distinguishing quantum and classical physics \cite{cirac2012goals}. However, it is clear that entanglement alone does not fully capture the quantum nature of a state \cite{gottesman1997stabilizer,aaronson2004improved}. For instance, stabilizer states, those generated by Clifford operations CNOT, Hadamard, and phase gates, can exhibit significant entanglement while remaining efficiently classically simulable, as guaranteed by the Gottesman-Knill theorem \cite{gottesman1997stabilizer}. The degree to which a quantum state deviates from the stabilizer set is known as \emph{non-stabilizerness} or \emph{magic}, and it is now understood to be a critical resource for quantum computational advantage.

In this numerical study, we investigate the magic content of quantum field theory ground states. Recent years have seen growing interest in simulating quantum field theories and lattice gauge theories using quantum devices \cite{banuls2020simulating,preskill2018quantum,bernien2017probing}. A question in this context is to determine physical features of a quantum field theory that pose barriers to classical simulation. As a concrete example, in this work we study the $(1+1)$-dimensional Schwinger model, quantum electrodynamics in one spatial and one temporal dimension, in the presence of a topological theta term. Despite its apparent simplicity, this model exhibits confinement and screening thereby serving as a qualitative toy model of QCD; it has become a benchmark for quantum simulation of gauge theories \cite{muschik2017u,klco2018quantum,chakraborty2001digital,de2022quantum}.

To quantify non-stabilizerness, a variety of measures have been proposed, including Wigner function negativity, Wigner entropy, and various distance-based metrics \cite{gross2007non,veitch2012negative,howard2017application,liu2022many}. In this work, we focus on the stabilizer Rényi entropy \cite{leone2022stabilizer}, which enjoys a variety of classical and quantum algorithms for its computation \cite{stratton2025algorithm,liu2025nonequilibrium,lami2023nonstabilizerness,haug2024efficient,ding2025evaluating,tarabunga2023many}. In particular, we apply a method recently introduced in \cite{sinibaldi2025non}, where the stabilizer Rényi entropy is computed from a variational neural network representation of the ground state.

Neural network quantum states (NNQS) \cite{carleo2017solving} are a variational ansatz for many-body wavefunctions, with successes in simulating strongly correlated systems (see \cite{lange2024architectures} for a review). Their application to quantum field theory models is relatively recent. Our results suggest that NNQS methods can capture not only the energy of ground states in this context but also their computational non-classicality, as measured by stabilizer R\'enyi entropy. 

The paper is organised as follows. In section \ref{sec:prelim} we review the basic aspects of the neural network quantum state (NNQS) ansatz and stabilizer R\'enyi entropy (SRE). We then discuss how SRE may be estimated using a NNQS parametrisation of a state. In section \ref{sec:schwinger} we discuss the physical context of the work and discuss the lattice regularisation of the Schwinger model. Finally, in section \ref{sec:results} we present numerical results for the SRE of the Schwinger model.

\section{Preliminaries}\label{sec:prelim}
In this section we review the basics of neural network quantum states and stabilizer theory. We consider a Hilbert space of qubits $\mathcal{H} = (\mathbb{C}^2)^{\otimes N}$ and write the computational basis states as $|x \rangle$ with $x \in \mathbb{Z}_2^N$. We consider a local Hamiltonian $H$ with a non-degenerate ground state denoted $|\Omega \rangle$. As an example, in this work we later consider the following spin Hamiltonian on $N$ qubits expressed in terms of Pauli matrices $X_n$, $Y_n$ and $Z_n$ acting on the n\textsuperscript{th} qubit as follows
\begin{equation}
\begin{split}
    H &= J \sum_{n=0}^{N-2} \left(\sum_{i=0}^n \frac{Z_i+(-1)^i}{2} + \frac{\vartheta_n}{2\pi}\right)^2\\ &+ \frac{w}{2} \sum_{n=0}^{N-2}(X_n X_{n+1} + Y_n Y_{n+1}) + 
    \frac{m}{2}\sum_{n=0}^{N-1}(-1)^n Z_n.
\end{split}    
\end{equation}

We discuss the quantum field theory origin of this model and the physical origin of the parameters $\vartheta$, $w$ and $m$ in the following section.

\subsection{Neural network quantum states}
\begin{figure}
    \centering
    \includegraphics[scale=0.18]{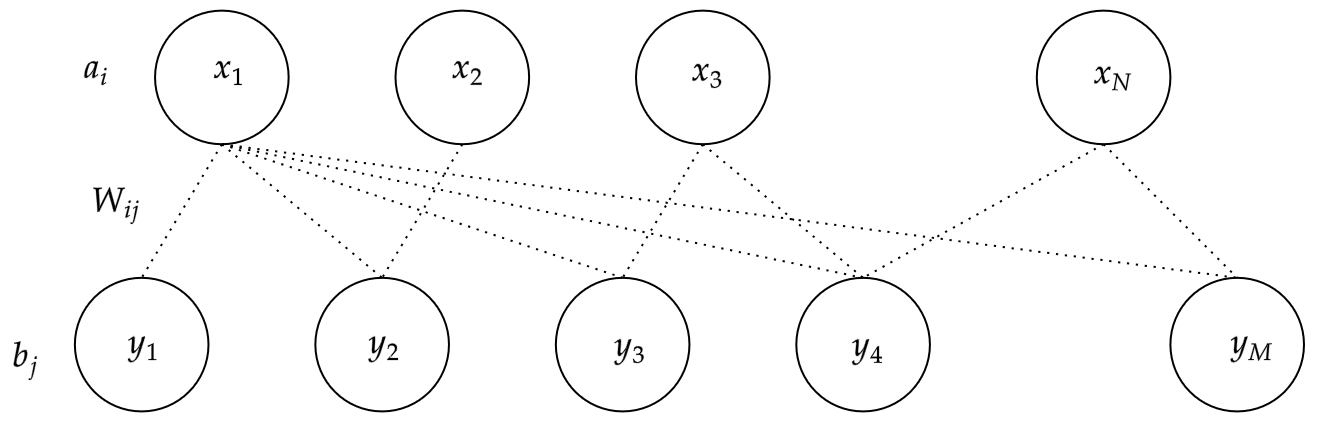}
  \caption{The RBM architecture $\psi_\mathbf{W}$ with $N$ binary visible units $x \in \mathbb{Z}_2^N$ and $M$ binary hidden units $y \in \mathbb{Z}_2^M$. Associated weights and biases are denoted $\mathbf{W} = (W_{ij},a_i,b_j)$.}
  \label{fig:RBM}
\end{figure}

Quantum many-body ground states may be encoded as neural network states \cite{carleo2017solving}. We consider a complex valued restricted Boltzmann machine (RBM) architecture with binary inputs, a visible layer of width $N$ and a single hidden layer of width $M$. We consider corresponding complex weights $\mathbf{W} = (W,a,b)$ where $W$ is an $N \times M$ weight matrix and $a$ and $b$ are complex bias vectors of length $N$ and $M$ respectively. Restricted means that the weight matrix $W$ has no inter-layer connections---this network architecture is illustrated in figure \ref{fig:RBM}. The RBM is a complex valued function $\psi_{\mathbf{W}}: \mathbb{Z}_2^N \to \mathbb{C}$ defined as follows. We first define an energy function 
\begin{equation}
\mathcal{E}(x,y;\mathbf{W}) = \sum_{i=1}^N \sum_{j=1}^M x_i W_{ij} y_j + \sum_{i=1}^M a_i x_i + \sum_{j=1}^M b_j y_j,
\end{equation}
in terms of which the (unnormalised) RBM is given by
\begin{equation}\label{eq:summedRBM}
\begin{split}
    \psi_{\mathbf{W}}(x) &= \sum_{y \in \mathbb{Z}_2^M}e^{-\mathcal{E}(x,y;\theta)} \\
    &= e^{\sum_{i=1}^N x_i a_i} \prod_{j=1}^M 2\cosh(\sum_{i=1}^N W_{ij}x_i + b_j).
\end{split}    
\end{equation}
In the second line we use the restricted property of the Boltzmann machine to perform the sum over hidden units explicitly.

On the other hand, the wave function of a (possibly unnormalised) state $|\psi \rangle$ in a system of qubits $\mathcal{H}$, denoted $\psi(x) = \langle x |\psi \rangle$, similarly defines a complex function $\psi: \mathbb{Z}_2^N \to \mathbb{C}$. A neural network quantum state uses the RBM architecture as a variational ansatz for the wave function and encodes $\langle \psi | x \rangle = \psi_{\mathbf{W}}(x)$.

\subsubsection*{Training}
We have discussed above the encoding of the state $|\psi\rangle$ as a Boltzmann machine $\psi_{\mathbf{W}}: \mathbb{Z}_2^N \to \mathbb{C}$. We now discuss now to optimise the weights $\mathbf{W}$ so as to minimise the energy of the state $|\psi\rangle$ and thereby approximate the ground state. To this end, we consider a loss function given by the quantum expected value of the energy 
\begin{equation}
    l(\mathbf{W}) := \langle H \rangle_{\mathbf{W}} = \frac{\langle \Omega | H | \Omega \rangle}{\langle \Omega | \Omega \rangle}.
\end{equation}
A useful observation is that the quantum expectation value may be re-expressed as a \textit{classical} expectation value. We first observe that the complex RBM defines a parametrised probability distribution on $\mathbb{Z}_2^N$ by
\begin{equation}
    p_{\mathbf{W}}(x) = \frac{|\psi_\mathbf{W}(x)|^2}{\sum_{x} |\psi_{\mathbf{W}}(x)|^2}.
\end{equation}
and the loss function may be expressed as an expectation value over this distribution 
\begin{equation}
    \langle H \rangle_{\mathbf{W}} = \mathbb{E}_{x \sim p_{\mathbf{W}}}\left[E_{\text{loc.}}(x)\right],
\end{equation}
where $E_{\text{loc.}}(x)$ denotes the local energy defined by 
\begin{equation}
    E_{\text{loc.}}(x) = \sum_{x' \in \mathbb{Z}_2^N} H_{x,x'}\frac{\psi_{\mathbf{W}}(x')}{\psi_{\mathbf{W}}(x)}.
\end{equation}
The terms $H_{x,x'} = \langle x | H |x'\rangle$ are the matrix elements of the Hamiltonian in the computational basis \footnote{Note that for a sufficiently local Hamiltonian this term is efficiently computable, i.e. the naively exponential sum actually only has polynomially many non-zero terms.}. The gradients of the loss function are similarly expressible as expectation values over $p_{\mathbf{W}}$:
\begin{equation}
\begin{split}
    &\partial_{\mathbf{W}_k} \mathbb{E}\left[E_{\text{loc.}}(x)\right] =  \\&\mathbb{E}\left[\frac{\partial_{\mathbf{W}_k} \psi_{\mathbf{W}}(x)}{ \psi_{\mathbf{W}}(x)} E_{\text{loc.}}(x)\right] - \mathbb{E}[E_{\text{loc.}}(x)]\mathbb{E}\left[\frac{\partial_{\mathbf{W}_k} \psi_{\mathbf{W}}(x)}{ \psi_{\mathbf{W}}(x)}\right].
\end{split}    
\end{equation}
The loss function $l(\mathbf{W})$ may then be optimised by stochastic gradient descent (with a learning rate $\eta$) since the gradient updates
are efficient to evaluate, in particular closed form expressions of the derivatives follow immediately from equation \eqref{eq:summedRBM}. For further implementation details we refer the reader to \cite{vicentini2022netket}.

\subsection{Stabilizer R\'enyi entropy}
We now briefly review the definition of the stabilizer R\'enyi entropy of a qubit state. We refer the reader to \cite{leone2022stabilizer} for a more thorough treatment. 

We define an $N$-qubit Pauli string $\sigma$ as an element of $\{I,X,Y,Z\}^{\otimes N}$ where $X$, $Y$ and $Z$ denote the standard Pauli matrices. We write $P_N$ for the set of $N$-qubit Pauli strings and $\mathcal{P}_N = \{\pm 1, \pm i\} \times P_N$ for the corresponding Pauli group. The Clifford group is defined as the normalizer of the Pauli group in the space of unitaries acting on $\mathcal{H}$
\begin{equation}
    \mathcal{C}_N = \{U \, : \, U \mathcal{P}_N U^{\dag} \subset \mathcal{P}_N\},
\end{equation}
and is generated by the Hadamard gate, CNOT gate and a phase gate---these are the so-called Clifford gates \cite{gottesman1997stabilizer}. A general state $|\psi\rangle \in \mathcal{H}$ is called a stabilizer state if it may be constructed from $|0\rangle^{\otimes N}$ acting with Clifford gates. Whilst having potentially large entanglement, the Gottesmann-Knill theorem demonstrates that these states offer no quantum advantage in the sense that they are classically efficiently preparable.

It is important to quantify the degree of non-stabilizerness of a state $|\psi\rangle$, the so-called magic \cite{bravyi2005universal}. One such quantity is the stabilizer R\'enyi entropy \cite{leone2022stabilizer} defined by
\begin{equation}
    M_{\alpha}(|\psi \rangle) := \frac{1}{1-\alpha} \log \frac{1}{2^N}\sum_{\sigma \in P_N} \left(\frac{\langle \psi | \sigma | \psi \rangle}{\langle \psi | \psi \rangle}\right)^{2\alpha},
\end{equation}
where the sum is over $4^N$ Pauli strings. Provided $\alpha \ge 2$ this quantity is a magic monotone \cite{leone2024stabilizer} and is thus a measure of non-stabilizerness.

\subsubsection*{Replicated estimator}
In the case of $\alpha = 2$, recent work has demonstrated that $M_2$ may be expressed as a classical expectation value by the so-called replicated estimator method \cite{sinibaldi2025non,tarabunga2024magic}. We now briefly review this approach. We consider the four replicated qubit Hilbert space where we write a general computational basis state as
\begin{equation}
    |\eta \rangle = |x^{1}\rangle \otimes |x^{2}\rangle \otimes|x^{3}\rangle \otimes|x^{4}\rangle 
\end{equation}
where $x^{(i)}$ for $i=1,2,3,4$ are each length $N$ binary strings. Associated to a state $|\psi \rangle \in \mathcal{H}$, we define the replicated state $|\Phi\rangle = |\psi^* \rangle  |\psi^* \rangle  |\psi^* \rangle |\psi \rangle$. We then define a unitary operator $\hat{U}$ on the replicated Hilbert space by
\begin{equation}
    \hat{U} | \eta \rangle = | x^2 \cdot x^3 \cdot x^4 \rangle | x^1 \cdot x^3 \cdot x^4 \rangle | x^1 \cdot x^2 \cdot x^4 \rangle | x^1 \cdot x^2 \cdot x^3 \rangle 
\end{equation}
where $x\cdot y$ denotes the binary XOR or Hadamard product between strings. In terms of these objects, one finds the following expression for the magic
\begin{align}\label{eq:replicatedestimator}
\exp\left(-M_2(|\psi \rangle)\right) 
&= \frac{\langle \Phi | \hat{U} | \Phi \rangle}{\langle \Phi | \Phi \rangle}
= \sum_{\eta} \frac{|\Phi(\eta)|^2}{\langle \Phi | \Phi \rangle}
\left[ \frac{\langle \eta | \hat{U} | \Phi \rangle}{\langle \eta | \Phi \rangle} \right] \notag \\
&= \mathbb{E}_{\eta \sim |\Phi(\eta)|^2} 
\left[ \frac{\langle \eta | \hat{U} | \Phi \rangle}{\langle \eta | \Phi \rangle} \right].
\end{align}
When the state $|\psi \rangle$ is encoded as an RBM, this quantity may be efficiently sampled. 

\section{The Schwinger model}\label{sec:schwinger}
In this work we study the stabilizer R\'enyi entropy of the Schwinger model \cite{schwinger1962gauge,coleman1975charge}, describing electrodynamics in $(1+1)$d spacetime. The Lagrangian consists of a Dirac fermion $\psi$ of mass $m$ coupled (with strength $g$) to a photon $A_\mu$ together with a theta term $\theta$, explicitly
\begin{equation}
    \mathcal{L} = -\frac{1}{4}F_{\mu \nu} F^{\mu \nu} + \frac{g\theta}{4 \pi} \epsilon_{\mu \nu}F^{\mu \nu} + i \bar{\psi}\gamma^{\mu}(\partial_{\mu} + i g A_{\mu})\psi - m \bar{\psi}\psi.
\end{equation}
The continuum Hamiltonian in the gauge $A_0 = 0$ reads
\begin{equation}
    H = \int \mathrm{d}x \,\frac{1}{2}\left(\Pi - \frac{g\theta}{2\pi}\right)^2 - i \bar{\psi} \gamma^1(\partial_1 + i gA_1)\psi + m\psi \bar{\psi}.
\end{equation}
The Kogut-Susskind approach \cite{kogut1975hamiltonian} regularises the theory as a spin system on a lattice of $N$ spins with lattice spacing $a$ describing the system in a physical volume $L = Na$. In this setting we may also introduce probe charges $+q$ and $-q$ at lattice sites $l_0$ and $l_0 + l$ respectively by promoting the theta term to a position dependent field:
\begin{equation}
\vartheta_n =
\begin{cases}
2\pi q + \theta_0, & \ell_0 \leq n < \ell_0 + \ell, \\
\theta_0, & \text{otherwise},
\end{cases}
\end{equation}
where $l_0 = (N-1-l)/2$.
The qubit regularisation is realised by introducing staggered fermions followed by a Jordan-Wigner transformation yielding the Hamiltonian
\begin{equation}
\begin{split}
    H &= J \sum_{n=0}^{N-2} \left(\sum_{i=0}^n \frac{Z_i+(-1)^i}{2} + \frac{\vartheta_n}{2\pi}\right)^2\\ &+ \frac{w}{2} \sum_{n=0}^{N-2}(X_n X_{n+1} + Y_n Y_{n+1}) + 
    \frac{m}{2}\sum_{n=0}^{N-1}(-1)^n Z_n.
\end{split}    
\end{equation}
The spin model parameters $J$ and $w$ are related to the field theory variables $a$ and $g$ via:
\begin{equation}
    J = \frac{g^2a}{2},\quad w = \frac{1}{2a}.
\end{equation}
The details of the derivation of this model are reviewed in more detail by, for example, Honda \textit{et. al.} \cite{honda2022classically}. 

An important motivation to study this model is that it exhibits qualitatively similar features to QCD; it has screening and confinement phases. The potential between two probe charges $\pm q$ exhibits different behaviour depending on the mass $m$ of the fermion and the value of $q$. For massless fermions ($m = 0$), all probe charges are screened. For massive fermions ($m \neq 0$), non-integer charges experience a confining potential that grows linearly with separation, while integer charges are screened. Indeed, at zero mass the potential in infinite volume is available in closed-form as
\begin{equation}\label{eq:potential_infinite}
V^{(0)}(\ell) = \frac{\sqrt{\pi} q^2 g}{2} \left( 1 - e^{- \frac{g \ell}{\sqrt{\pi}}} \right),
\end{equation}
and in finite volume one finds \cite{honda2022classically}
\begin{equation}\label{eq:potential_finite}
V_f^{(0)}(\ell) = \frac{\sqrt{\pi} q^2 g}{2} \cdot 
\frac{
\left(1 - e^{- \frac{g \ell}{\sqrt{\pi}}} \right)
\left(1 + e^{- \frac{g (L - \ell)}{\sqrt{\pi}}} \right)
}{
1 + e^{- \frac{g L}{\sqrt{\pi}}}
}.
\end{equation}

\section{Numerical results}\label{sec:results}

In this section we present our numerical results. We first verify the accuracy of the neural network ground state before turning to the stochastic sampling of the stabilizer R\'enyi entropy. Throughout this section, the numerical computations are implemented using the NetKet Python package \cite{vicentini2022netket}.

\subsection{Ground state}
We first evaluate the accuracy of the neural network encoding of the Schwinger model ground state. The neural network quantum state (NNQS) can be trained to high accuracy for simple spin models such as the transverse-field Ising model, using a relatively large learning rate (e.g. $\eta = 0.1$), a small number of hidden units $M=N$, and modest sample number for gradient estimation. In contrast, we found that training NNQS for the Schwinger model is more delicate. The optimisation process is prone to getting trapped in local minima, necessitating a smaller learning rate and weight perturbations \footnote{We perturb the weights $W_{ij}$ uniformly scaled to $0.1\%$ of the maximum when the variance of neural network quantum state is near zero, around $\sigma^2 = 0.001$)}. 

The NNQS is trained initially for $l = 0$ and $q=0$ with hidden layer width $M = N$, $K = 10^5$ Monte Carlo samples, a learning rate of $\eta = 0.001$, and $5000$ iterations. The model is then successively trained for $l = 1, 2, \ldots, l_{\text{Max.}}$, using the trained NNQS from the preceding $l$ as the initialization for the next. For $l>0$ we use a larger learning rate, fewer samples and small iteration size. In the following numerical simulations, each round of training for a given set of parameters takes 1hr and 2hrs for \(N=15\) and \(N=21\) respectively, using the A100 GPU available in Google Colab. 

Firstly, we compute the neural network ground state energies under the trained weights
\begin{equation}
    E_g(\theta_0,q,l) := \langle \psi_{\mathbf{W}} | H | \psi_{\mathbf{W}}\rangle = \mathbb{E}_{x \sim p_{\mathbf{W}}}\left[E_{\text{loc.}}(x)\right].
\end{equation}
When the site number is low ($N=15$) we evaluate the expectation value on the right hand side exactly. Table \ref{tab:energy_comparison} compares these values with the exact diagonalisation ground state energies computed using the Lanczos algorithm.

\begin{figure}
    \begin{subfigure}{1\linewidth}
    \centering
        \includegraphics[width=1\linewidth]{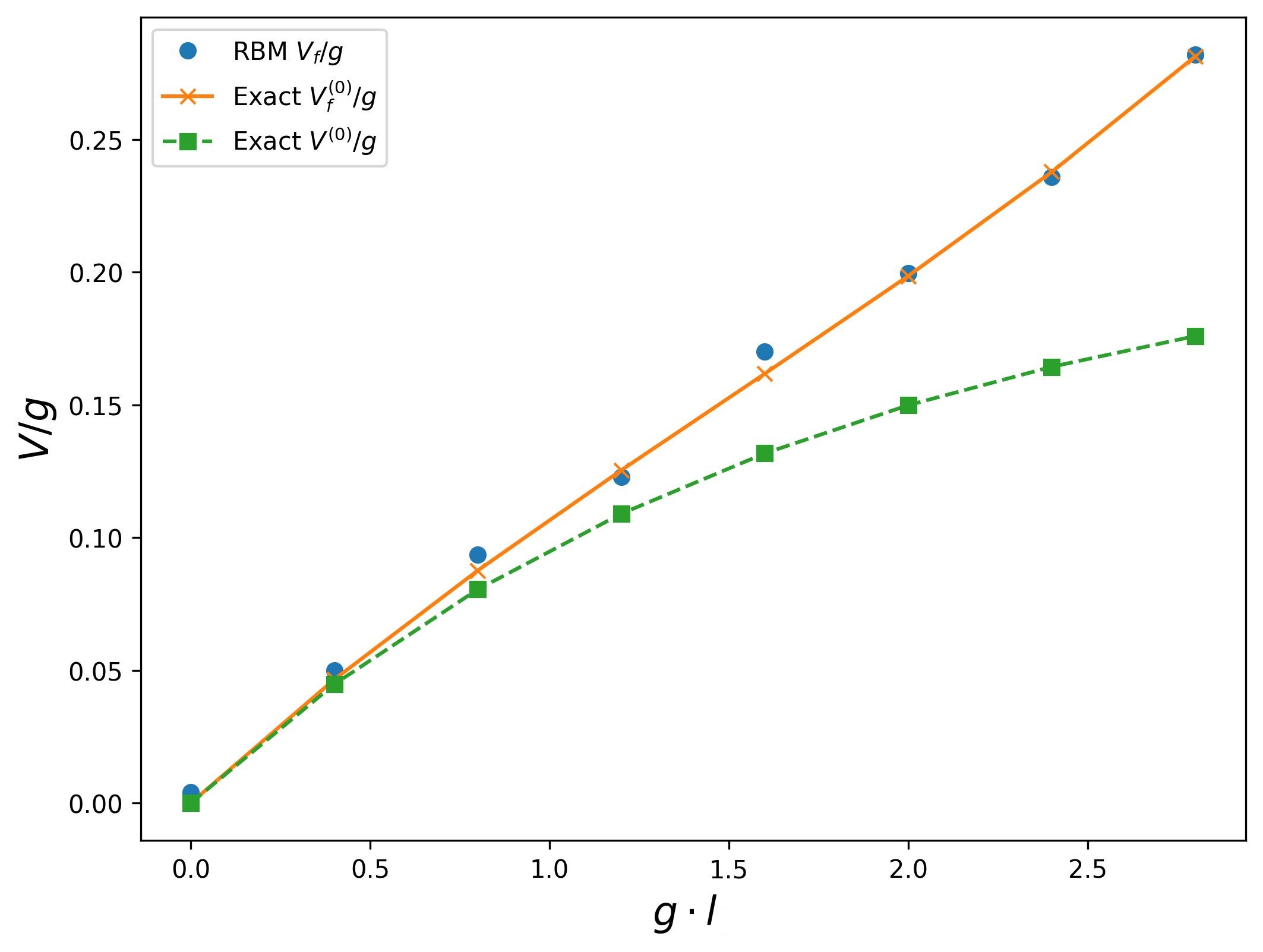}     
    \end{subfigure}
    \begin{subfigure}{1\linewidth}
    \centering
        \includegraphics[width=1\linewidth]{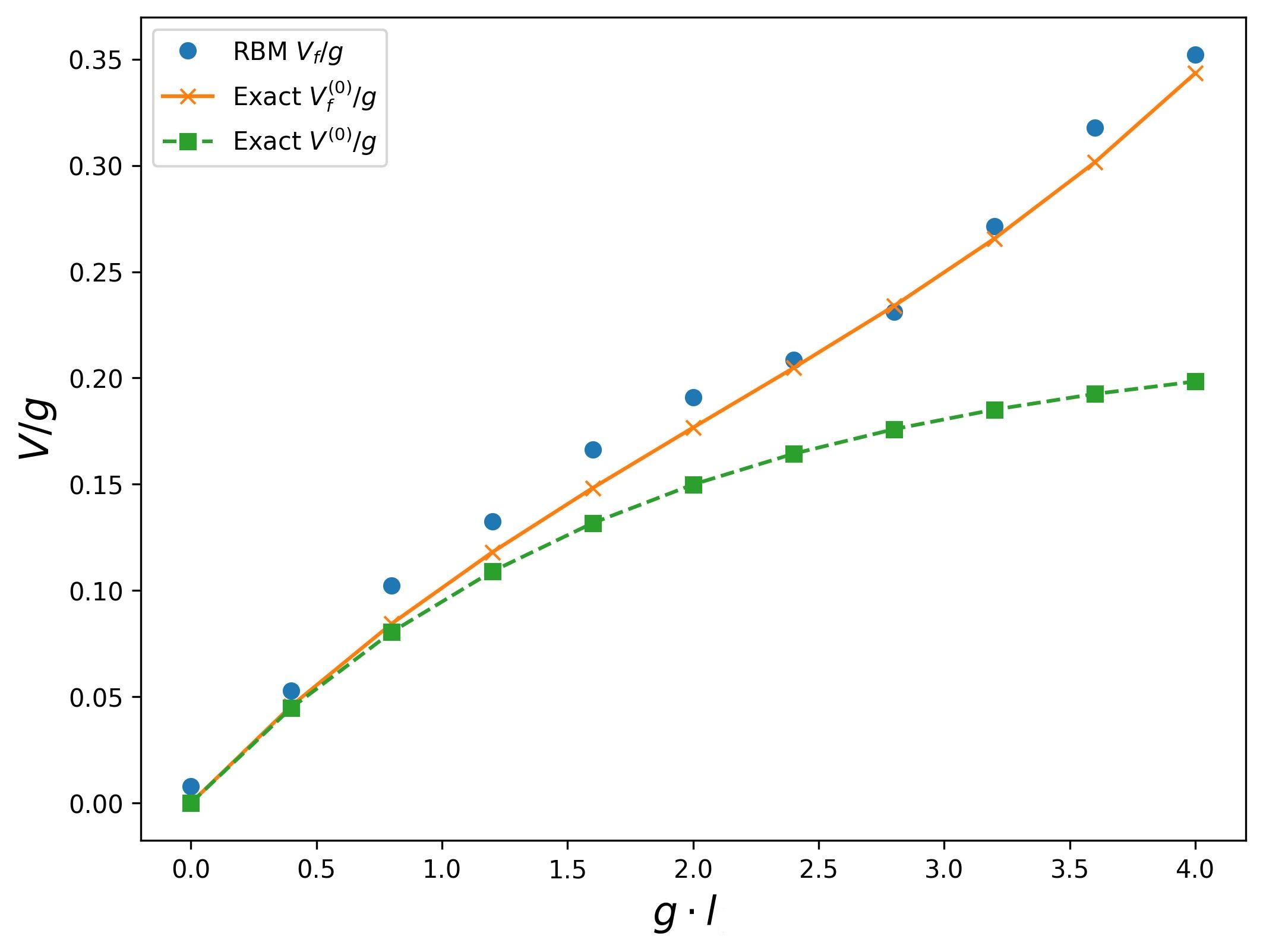}     
    \end{subfigure}
    \begin{subfigure}{1\linewidth}
    \centering
        \includegraphics[width=1\linewidth]{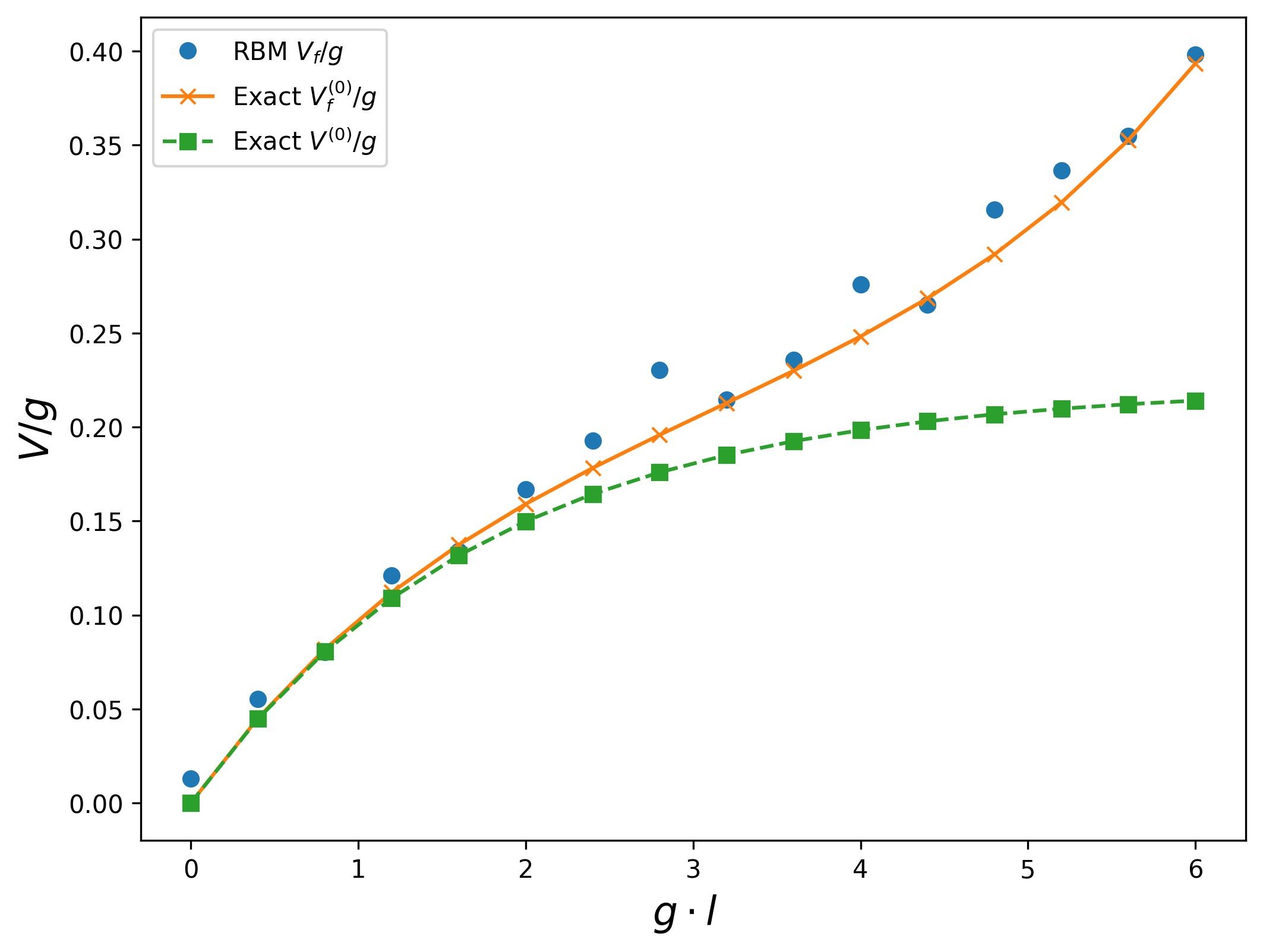}     
    \end{subfigure}
    \caption{Comparison of the NNQS numerical potential \eqref{eq:potential_numerical} with the finite volume exact potential \eqref{eq:potential_finite} and the infinite volume exact potential \eqref{eq:potential_infinite} for $N=15$, $N=21$ and $N=31$ lattice sites. Model parameters are $g=0.2$, $a=1$, $m=0$, $q=0.5$, $\theta_0=0$.}
    \label{fig:potential}
\end{figure}

Secondly, we examine the potential energy of the model as a function of the separation between probe charges, defined by
\begin{equation}\label{eq:potential_numerical}
V_f(l) = E_g(\theta_0, q, l) - E_g(0,0,0).
\end{equation}
In Figure~\ref{fig:potential}, we compare the results obtained from the neural network ground state with those from exact diagonalisation.

\begin{table}[htbp]
  \centering
  \small  % Reduce font size to fit in one column
  \renewcommand{\arraystretch}{1.2}
  \begin{tabular}{@{}l r@{}}
    \toprule
    \textbf{$g \cdot l$} & \textbf{Distance $d$ (\%)} \\
    \midrule
    0.0 & 0.041539 \\
    0.8 & 0.022865 \\
    1.6 & 0.064646 \\
    2.4 & 0.025786 \\
    3.2 & 0.027567 \\
    4.0 & 0.018546 \\
    4.8 & 0.021070 \\
    5.6 & 0.023642 \\
    \bottomrule
  \end{tabular}
  \caption{Distance $d$ between $N=15$ neural network and exact diagonalisation states for \(g=0.4,\,a=1,\,m=0,\,q=0.5,\,\theta_0=0\) with \(l = 0,\,2,\,4,\ldots,\,14\).}
  \label{tab:d}
\end{table}

Finally, we evaluate the trace distance between the neural network quantum state and the exact diagonalisation state $|\Omega \rangle$:
\begin{equation}
    d\equiv 1 - \left| \langle \Omega|\psi_{\mathbf{W}}\rangle\right|^2,
\end{equation}
this metric is presented in Table~\ref{tab:d}. We find that the maximum trace distance is less than $0.05\%$.

\begin{figure}
    \begin{subfigure}{1\linewidth}
    \centering
        \includegraphics[width=1\linewidth]{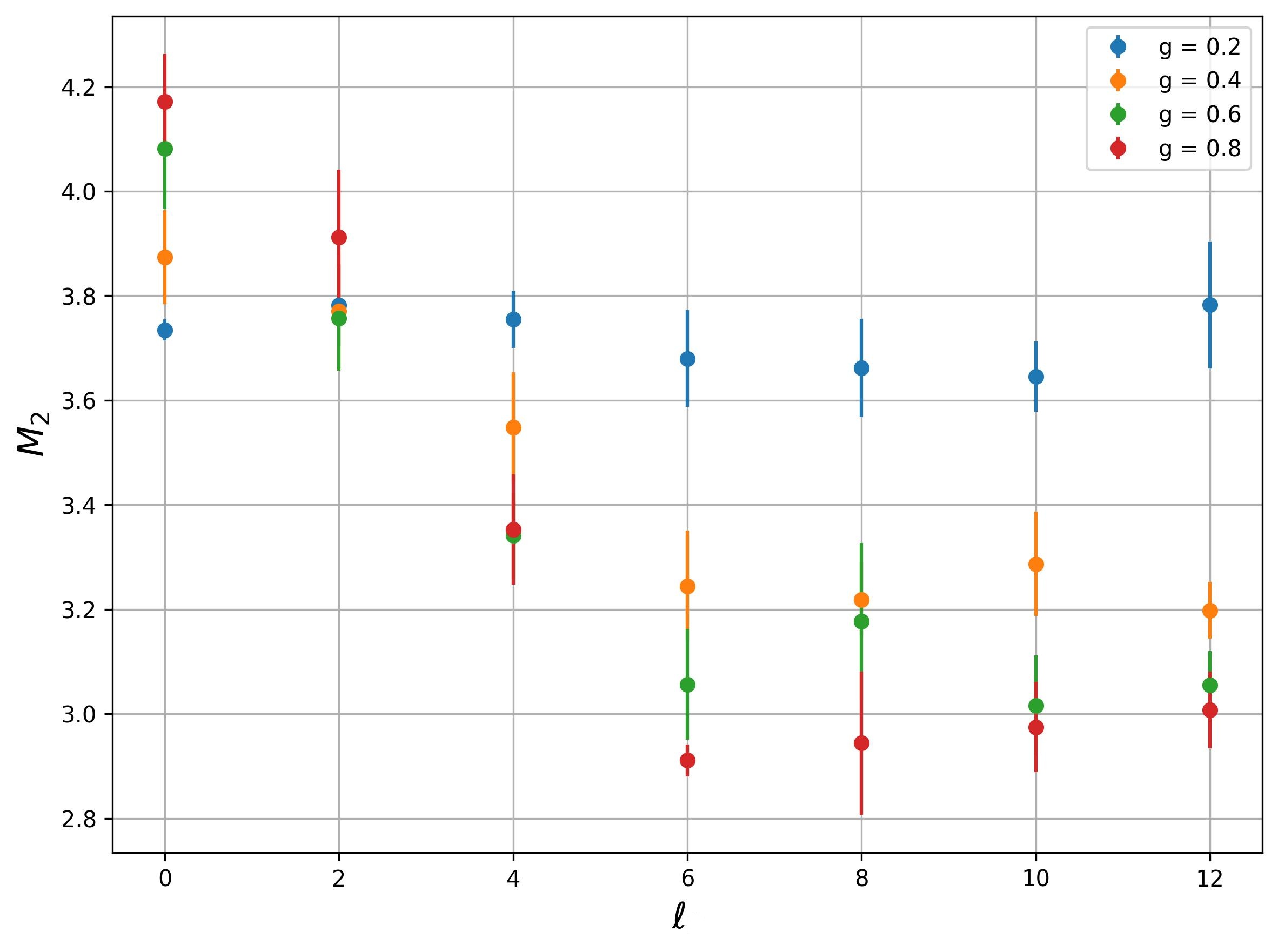}     
    \end{subfigure}
    \begin{subfigure}{1\linewidth}
    \centering
        \includegraphics[width=1\linewidth]{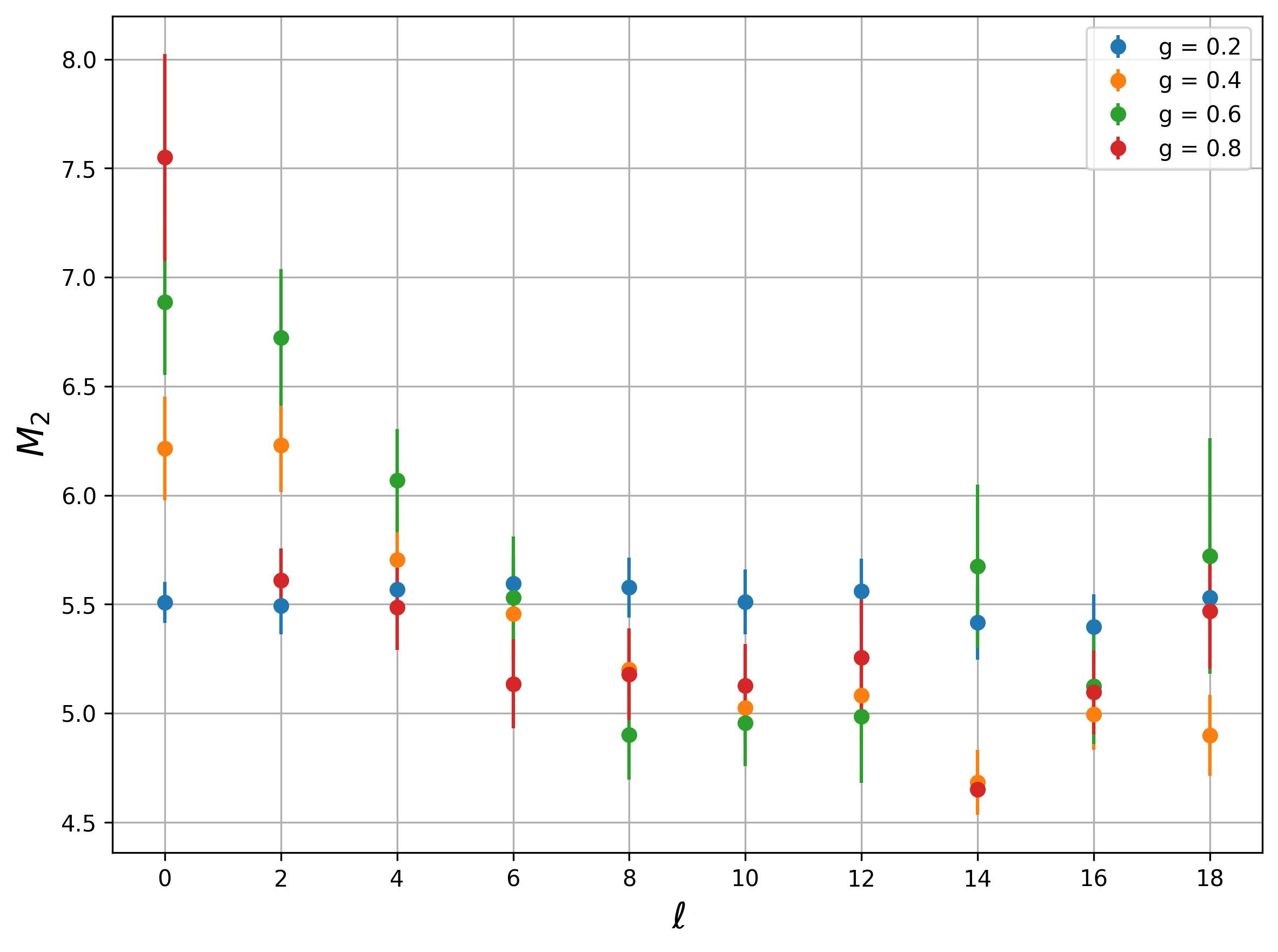}     
    \end{subfigure}
    \caption{Numerical stabilizer R\'enyi entropy versus $gl$ for $N=15$ and $N=21$ lattice sites in the massless theory. Parameters are $g \in \{0.2,0.4,0.6,0.8\}$, $a=1$, $m=0$, $q=0.5$, $\theta_0=0$.}
    \label{fig:magic1}
\end{figure}
\begin{figure}
    \begin{subfigure}{1\linewidth}
        \centering
        \includegraphics[width=1\linewidth]{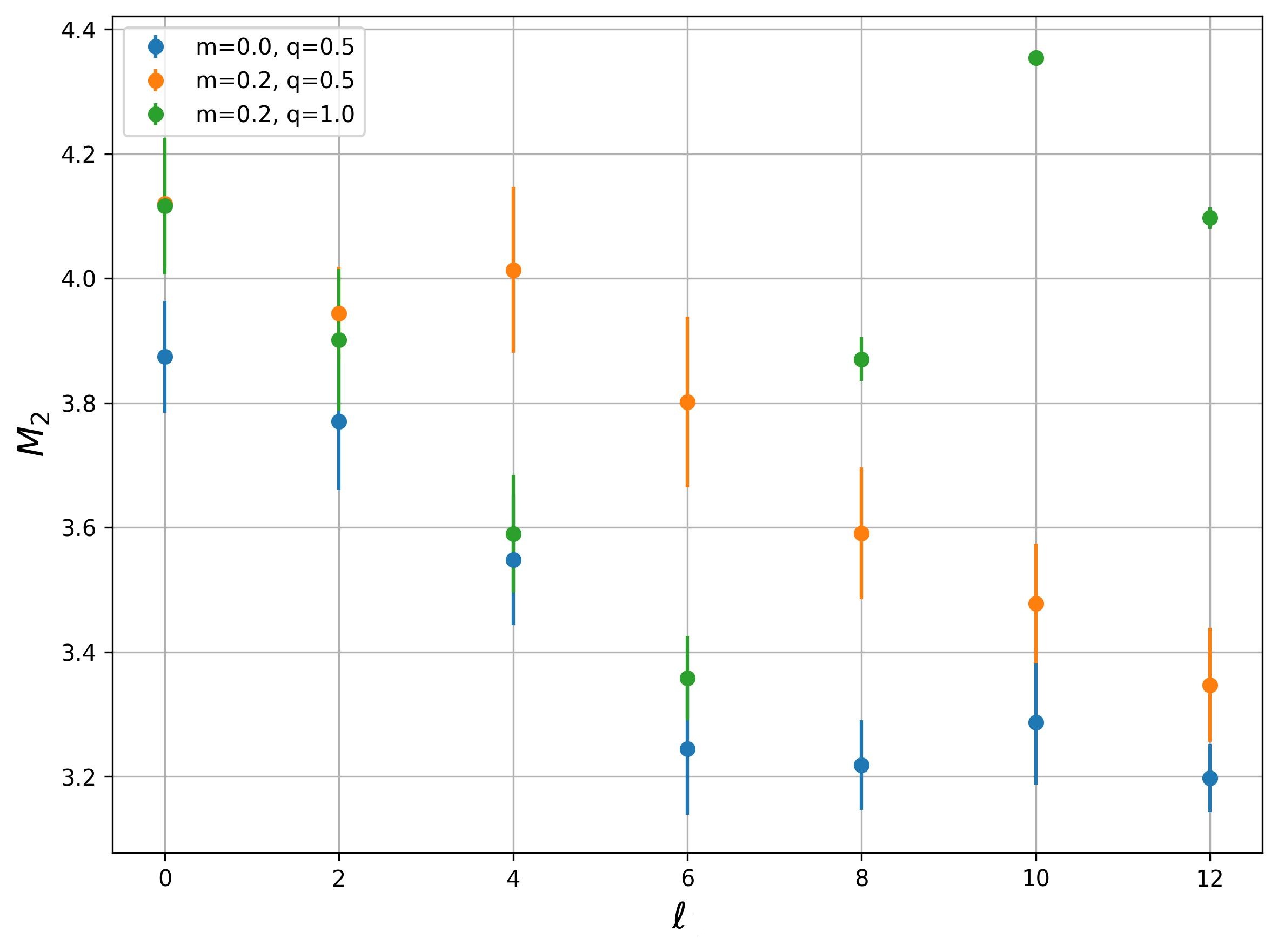}   
    \end{subfigure}
    \begin{subfigure}{1\linewidth}
        \centering
        \includegraphics[width=1\linewidth]{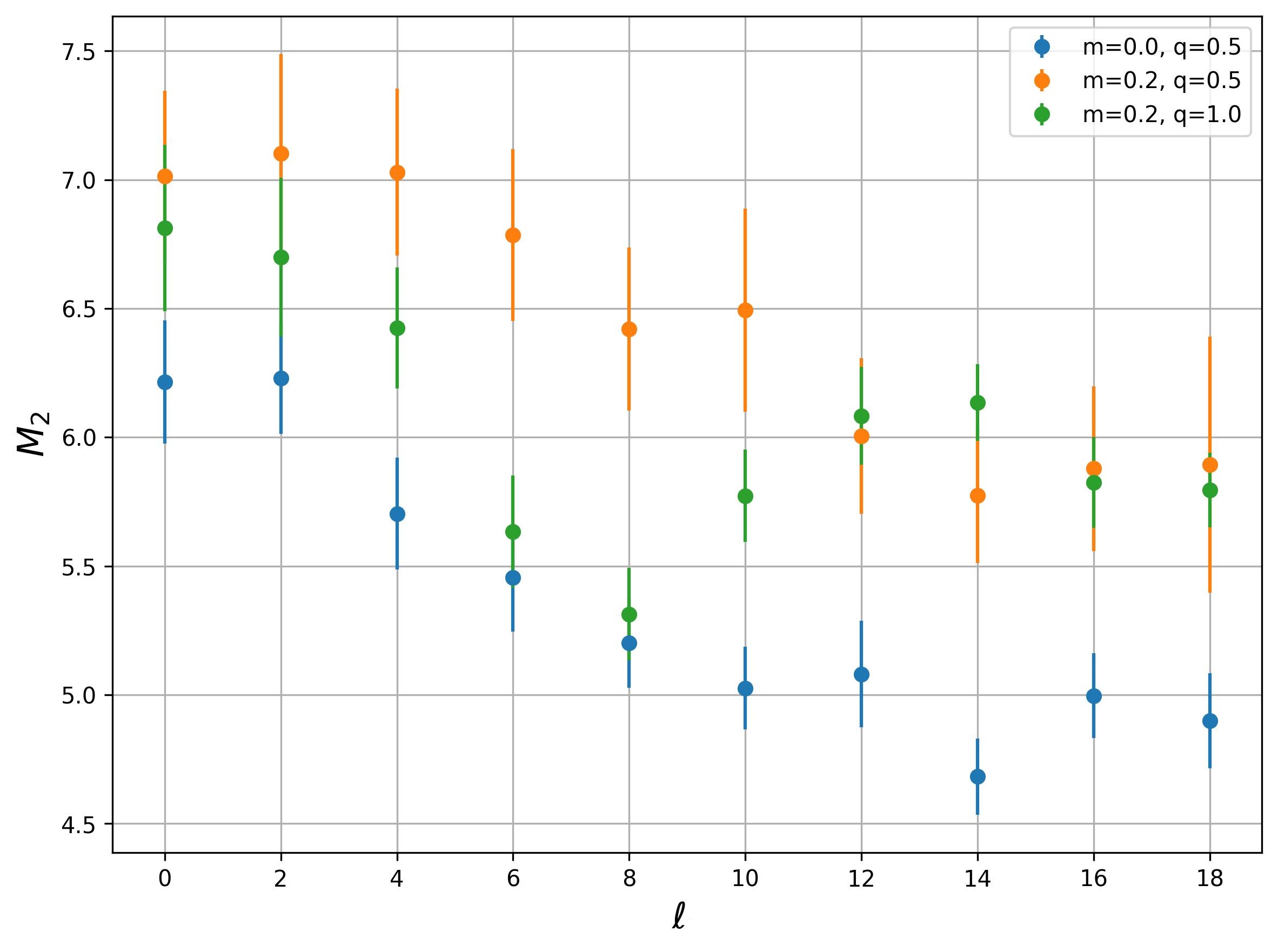}   
    \end{subfigure}
    \caption{Stabilizer R\'enyi entropy in confinement and screening phases with $N=15$ and $N=21$ lattice sites. Model parameters are $g=0.4$, $a=1$ and $\theta_0 = 0$.}
    \label{fig:magic3}
\end{figure}

\subsection{Stabilizer Entropy}
We now turn to the computation of the stabilizer R\'enyi entropy. We train the neural network ground state for $N=15$ and $N=21$ sites for $l=1,\ldots,N_{\text{Max.}}$ for a range of $g$ with $q=0.5$. As discussed in the previous section the stabilizer R\'enyi entropy may be expressed as 
\begin{equation}
    \exp(-M_2(|\psi\rangle)) = \mathbb{E}_{\eta \sim |\Psi(\eta)|^2}\left[ \frac{\langle \boldsymbol{\eta} | \hat{U} | \Psi \rangle}{\langle \Psi | \Psi \rangle}\right],
\end{equation}
and the expectation value on the right hand side may be estimated by Monte Carlo sampling our neural network parametrisation of the ground state. The replicated estimator sampling \eqref{eq:replicatedestimator} takes 30s per round of $10^6$ sampling using the A100 GPU. In the massless case, the results are shown in Fig.~\ref{fig:magic1} for $N=15$ and $N=21$ sites. We then consider the massive case. We fix a representative coupling $g=0.4$ and compute the SRE with example integer and half-integer charges. In all plots we include empirical $95\%$ confidence intervals.

\begin{table*}[htbp]
  \centering
  \renewcommand{\arraystretch}{1.4}  % Increases row height for readability
  \begin{tabularx}{\textwidth}{@{} >{\raggedright\arraybackslash}X 
                                   >{\raggedright\arraybackslash}X 
                                   >{\raggedright\arraybackslash}X @{}}
    \toprule
    \textbf{$gl$} & \textbf{NNQS $E_0/g$} & \textbf{ED $E_0/g$} \\
    \midrule
    0.0
      & -16.300719
      & -16.295373\\
    
    0.8
      & -16.218140
      & -16.212814\\

    1.6
      & -16.149031
      & -16.150611\\
    
    2.4
      & -16.104108
      & -16.105701\\
      
    3.2
      & -16.052260
      & -16.059035\\
      
    4.0
      & -16.029920
      & -16.025538\\

    4.8
      & -15.967276
      & -15.969200\\

    5.6
      & -15.916560
      & -15.969200\\
    \bottomrule
  \end{tabularx}
  \caption{Comparison between neural network and exact diagonalisation computation of the ground state energy of the $N=15$ site massless theory as $l$ varies. Parameters are $g=0.4$, $a=1$, $m=0$, $q=0.5$ and $\theta_0=0$.} 
  \label{tab:energy_comparison}
\end{table*}

\section{Discussion}
In this work, we have demonstrated that neural network quantum states (NNQS) are an effective variational method for studying the magic properties of quantum field theory ground states, specifically within the qubit-regularised Schwinger model. By using a recently proposed replicated estimator method, we computed the stabilizer R\'enyi entropy, a stabilizer monotone, across a range of system parameters. 

Our results show that NNQS not only accurately approximate the energy of lattice gauge theory ground states, but also capture features related to the classical hardness of these states when regularised as a system of qubits. We observe in both figures \ref{fig:magic1} and \ref{fig:magic3} that non-stabilizerness increases as external probe charges are brought closer together. Our numerical results also suggest a qualitatively different behaviour between the screening and confinement phases. This behaviour is more pronounced at strong coupling $g$. In figure \ref{fig:magic3} we note that in the screening phase the stabilizer R\'enyi entropy has an apparent local minimum when the charges are maximally separated from each other and the boundary whereas in the confinement phase the stabilizer R\'enyi entropy is monotonically decreasing.  

From a physical perspective, our results provide an initial step toward using quantum magic as a diagnostic of phase transitions in quantum field theory. In future work, it would be valuable to develop an analytic understanding of this phenomenon in the Schwinger model, clarifying the origin of the distinct behaviours observed in the screening and confinement phases.

\section*{Acknowledgements}
This work is supported by National Science and Technology Council of Taiwan under Grant No. NSTC 113-2112-M- 007-019. The numerical computations were carried out using Google Colaboratory (Colab), whose free GPU resources facilitated rapid prototyping and simulation.
The authors thank Po-Yao Chang, Masazumi Honda, Ying-Lin Li, Marco Paini and Cris Salvi for many interesting discussions and acknowledge \texttt{ChatGPT} by OpenAI for assistance in code development and debugging.

\bibliography{schwingermagic}% Produces the bibliography via BibTeX.

\end{document}